\documentclass[12pt]{iopart}
\usepackage{iopams}
\usepackage{dsfont}
\usepackage[dvips]{graphicx}
\usepackage{setspace}
\newcommand{\half}{\mbox{$\textstyle \frac{1}{2}$}}

\newcommand{\re}{\mbox{$\rm e$}}

\newcommand{\rd}{\mbox{$\rm d$}}

\begin{document}
\title[Coquaternionic mechanics]{On complexified mechanics and coquaternions}

\author[D~C~Brody \& E~M~Graefe]{Dorje~C~Brody and Eva-Maria Graefe}

\address{Department of Mathematics, Imperial College London,
London SW7 2AZ, UK}

\begin{abstract}
While real Hamiltonian mechanics and Hermitian quantum mechanics can 
both be cast in the framework of complex canonical equations, their complex 
generalisations have hitherto remained tangential. In this paper 
quaternionic and coquaternionic (split-signature analogue of quaternions) 
extensions of Hamiltonian mechanics are introduced, and are shown to offer 
a unifying framework for complexified classical and quantum mechanics. 
In particular, quantum theories characterised by complex Hamiltonians invariant 
under space-time reflection are shown to be equivalent to certain coquaternionic 
extensions of Hermitian quantum theories. One of the interesting consequences 
is that the space-time dimension of these systems is six, not four, on account 
of the structures of coquaternionic quantum mechanics. 
\end{abstract}
\vspace{0.4cm}


This paper concerns the relation between complexified classical 
and non-Hermitian quantum mechanics, and their surprising links to 
quaternionic and coquaternionic mechanics. The main finding is that 
complexified mechanical systems with real energies studied extensively 
in the literature over the past decade 
can alternatively be thought of as certain coquaternionic extensions 
of the underlying real mechanical systems. This identification leads 
to the possibility of employing algebraic techniques of quaternions 
and coquaternions to tackle some of the challenging open issues in 
complexified classical and quantum mechanics.  

Complex (i.e. non-Hermitian) Hamiltonians have long been employed to 
describe open quantum systems, decay and scattering phenomena 
\cite{Moiseyevbook}. Further, since the realisation that complex operators 
respecting space-time reflection (PT) symmetry may possess entirely real 
spectra~\cite{Bender}, there have been considerable research interests in 
examining both physical and mathematical properties of quantum systems 
described by non-Hermitian Hamiltonians with real spectra. More recently, 
the interest in these systems has increased notably, in part owing to 
experimental realisations of the phenomenon of the PT phase 
transition and other theoretically predicted effects~\cite{Guo,Zhao}. 

Complexified classical mechanics has also been studied intensely both 
in the context of semiclassical calculations and as a classical analogue 
of non-Hermitian quantum mechanics \cite{Xavi,kaushal,Bender2}. At 
classical level, complex-extended mechanics is characterised by a phase 
space spanned by complex canonical variables. The quantum analogue of 
a phase space---in the strict sense of a symplectic manifold upon which 
the dynamics is governed by Hamilton's equations of motion---is the 
space of pure states, i.e. the space of rays through the origin of the 
Hilbert space. On this ray space, the wave equation of Schr\"odinger in 
Hermitian quantum mechanics can be expressed as Hamilton's 
equations of motion \cite{strocchi,weinberg}. An apparent puzzle in this 
context is that the analysis of \textit{complex} extensions of quantum 
mechanics has thus far been confined to the \textit{real} state space 
\cite{BBJM,Mostafa2,Gunther} (with the exception of 
\cite{Nesterov}). Therefore, the nature of complexification 
considered so far for quantum mechanics has a rather different 
characteristic from that for classical mechanics. 
If classical phase spaces can be extended into the complex 
domain, it seems paradoxical that quantum state space cannot be 
extended in an analogous manner. (Quaternionic 
quantum mechanics~\cite{Finkelstein,Adler} can be 
viewed as representing `complex' extensions of the underlying real 
mechanics for real (Hermitian) quantum systems, although this is not 
the point of view commonly adopted.) 

In the present paper we propose a new approach to these 
apparently different complex generalisations by making use of 
quaternionic and coquaternionic formulations. The key idea is that 
phase-space variables consist of canonical conjugate pairs; 
hence a complexified mechanical system necessarily involves 
pairs of complex numbers. On the other hand, pairs of complex 
numbers can usefully be treated as a single quaternion. It is natural 
therefore to enquire whether complexified mechanics can be 
represented more concisely in terms of quaternions, inasmuch as 
real phase space variables can more conveniently be represented 
as a single complex number. A deeper 
motivation for such a study arises from 
the fact that symmetry properties of quaternions are closely related 
to symmetries of (Euclidean) space-time~\cite{Conway}; while 
complexified dynamical systems discussed here are characterised 
by Hamiltonians invariant under space-time reflection (PT) symmetry. 
The appearance of space-time symmetry in both complexified 
mechanics and quaternionic algebra suggests that the latter might 
constitute a useful tool for the investigation of the former. A further 
motivation comes from the observation that if the parity 
operator is assumed trace free, then the real dimensionality of 
the Hilbert space of a PT-symmetric quantum system is a multiple 
of four, not two~\cite{BBHM}. This suggests that quaternions \textit{a 
priori} are suitable candidates for the characterisation of phase 
space variables for such theories. 

With these observations in mind, we shall explore the possibility 
that complexified mechanics can alternatively be viewed as a 
version of quaternionic mechanics. We shall find, perhaps somewhat 
unexpectedly, that PT-symmetric systems are in fact related to 
coquaternions, rather than quaternions. In this connection it is worth 
remarking that symmetries of coquaternions are related to the Lorentz 
group, rather than Euclidean group, in the sense that every rotation in 
the Minkowski three-space can be expressed in terms of coquaternions. 
Hence PT symmetry \textit{a priori} has more in common with 
Lorentzian space-time than Euclidean space-time. It follows that 
PT-symmetric quantum mechanics, although related, are not equivalent 
to the traditional quaternionic quantum theories of \cite{Finkelstein,Adler}. 

Our hope is that the approach introduced here 
may ultimately help to tackle some open issues in the 
area of complexified mechanics that have only been 
answered at most partially. These include, among others: 
The classical-quantum correspondence of complexified mechanics \cite{Graefe,BHMW,Most06,Curt07}; the derivation of statistical 
mechanics of classical PT-symmetric systems~\cite{Jones}; the 
characterisation of combined systems (e.g., entanglement) in 
PT-symmetric quantum mechanics, and the formulation of a theory 
of space-time that is consistent with complex Hamiltonians.

\textit{Quaternions and coquaternions}. 
Although properties of quaternions are well known, we 
find it convenient to briefly summarise here some of the key 
relations. The algebra of quaternions involves three imaginary units, 
denoted here by $i$, $j$, $k$, satisfying the relation 
\begin{eqnarray}
i^2 = j^2 = k^2 = ijk = -1, \label{eq:1}
\end{eqnarray}
carved into the stone of the Brougham Bridge by Hamilton in an 
act of ``mathematical vandalism'' \cite{Baez}. It follows that 
quaternions satisfy the following cyclic relation:  
\begin{eqnarray}
ij=-ji=k, \quad jk=-kj=i, \quad ki=-ik=j.  \label{eq:2}
\end{eqnarray}
Hence matrix representations of the three imaginary 
units are just the Pauli matrices multiplied by the complex number.
 
A generic quaternion can be represented in the form 
$q = q_0 + i q_1 + j q_2 + k q_3$, where $\{q_i\}_{i=0,1,2,3}$ are all real. 
It is not difficult to see that 
quaternions that commute with all other quaternions are reals, and 
that the totality of quaternions that commute with a given quaternion 
forms a subset isomorphic to complex numbers. The conjugate of 
a quaternion $q$ is given by ${\bar q} = q_0 - i q_1 - j q_2 - k q_3$. 
It follows that the squared modulus of $q$ is 
${\bar q}q = q_0^2 + q_1^2 + q_2^2 + q_3^2$, and that the inverse 
of  $q$ is $q^{-1}={\bar q}/({\bar q}q)$. A quaternion can be 
expressed in polar form 
\begin{eqnarray}
q = |q| \re^{{\boldsymbol i}_q \theta_q} = |q|(\cos\theta_q + 
{\boldsymbol i}_q \sin\theta_q),
\end{eqnarray}
where 
${\boldsymbol i}_q = (iq_1+jq_2+kq_3)/|q|$ and 
$\theta_q = \cos^{-1}(q_0/|q|)$. 

The symmetry properties of quaternions are thus closely related to 
the group $SU(2)$. The $SU(1,1)$ analogues of quaternions 
are the coquaternions~\cite{Cockle}, also known as split 
quaternions. They satisfy the relation 
\begin{eqnarray}
i^2 = -1, \quad j^2 = k^2 = ijk = +1
\end{eqnarray}
in place of (\ref{eq:1}), and the skew-cyclic relation 
\begin{eqnarray}
ij=-ji=k, \quad jk=-kj=-i, \quad ki=-ik=j 
\end{eqnarray}
in place of (\ref{eq:2}). Like quaternions, the conjugate of a 
coquaternion $q = q_0 + i q_1 + j q_2 + k q_3$ is 
${\bar q} = q_0 - i q_1 - j q_2 - k q_3$. It follows that the 
squared modulus of a 
coquaternion is indefinite: ${\bar q}q = q_0^2 + q_1^2 - q_2^2 - 
q_3^2$. Unlike quaternions, a coquaternion need not have an 
inverse $q^{-1}={\bar q}/({\bar q}q)$ if it is null, i.e. if ${\bar q}q =0$. 
The polar decomposition of a coquaternion thus 
involves trigonometric and hyperbolic functions in the usual 
way. Unlike quaternions, however, coquaternions do not 
admit finite-dimensional unitary representations. 

\textit{Complex formulation of real classical and quantum dynamics}. 
Before we proceed to explore the quaternionic or coquaternionic 
extensions, it is useful to represent the dynamical theory 
of elementary mechanical systems in terms of complex phase-space 
variables~\cite{Mackey,wiener}. 
We start with a one-dimensional dynamical system, described by 
real canonical variables $p$ and $x$, and introduce the complex 
conjugate variables $z = (x + i p)/\sqrt{2}$ and ${\bar z} = (x - 
i p)/\sqrt{2}$. The canonical equation of motion then reads 
\begin{eqnarray}
i \frac{\rd z}{\rd t} = \frac{\partial H}{\partial {\bar z}}, \label{eq:7} 
\end{eqnarray}
where $H=H(z,{\bar z})$ denotes the Hamiltonian. In this formulation 
it becomes apparent in which way quantum Schr\"odinger dynamics 
can be viewed as a special case of Hamiltonian mechanics 
\cite{Mackey}. Consider a quantum system characterised more 
generally by a state vector $|z\rangle$, with components $z_n$. The 
Schr\"odinger equation, with a Hermitian Hamiltonian $\hat H$, can then 
be written in the form $i \dot z_{n}=\rd H /\rd {\bar z}_n$, where 
$H=\langle z| \hat H| z\rangle=\sum_{mn} H_{mn} {\bar z}_n z_m$ is 
the expectation value of $\hat H$. In particular, the Hamiltonian function 
of a quantum-mechanical system is bilinear in the canonical variables, 
that is, it is essentially the Hamiltonian of coupled harmonic oscillators. 

With this in mind, we shall devote a great part of the present 
paper to the investigation of the bilinear Hamiltonian 
$H={\bar z}z$, leading to the equation of motion 
$i \rd z/\rd t = z$.
The trajectories in the $z$-plane are concentric circles about the 
origin, traversed with a constant angular velocity, which define a 
linear critical point called a \textit{centre}. If we perform a Wick 
rotation, we obtain $\rd z/\rd t = z$ 
(or $\rd z/\rd t = -z$), which defines a linear critical 
point called a \textit{focus}. The trajectories are rays leaving (or 
approaching) the origin, traversed with radially increasing velocity. 
More generally, consider the equation $\rd z/\rd t = bz$, 
where $b$ is a complex constant. If both the real and imaginary 
parts of $b$ are nonzero, then the corresponding linear critical point 
is a \textit{vortex}, and the orbits consist of concentric spirals 
leaving (or approaching) the origin traversed with equal radially 
increasing velocities. Quantum mechanically, all fixed points of 
unitary dynamics are centers, whereas in a PT-symmetric quantum 
theory a pair of centres can turn into a pair of foci or vortices upon 
symmetry breaking. 

\textit{Complexified classical and quantum mechanics}. Let us 
now summarise briefly the conventional way in which classical and 
quantum mechanics are extended into the complex domain. The 
complexified classical mechanics typically starts from the real 
canonical equations of motion ${\dot p}=-\partial H/\partial x$ and 
${\dot x}=\partial H/\partial p$, and considers the complexification 
$p=p_0+i p_1$ and $x=x_0 + i x_1$. In terms 
of the four real variables $(p_0,x_0,p_1,x_1)$ the Hamiltonian 
now takes the form $H=H_0(p_0,x_0,p_1,x_1)+iH_1(p_0,x_0,p_1,x_1)$. 
Assuming that $H$ is analytic and thus satisfies the Cauchy-Riemann 
conditions $\partial H_0/\partial x_0=\partial H_1/\partial x_1$, 
$\partial H_0/\partial x_1=-\partial H_1/\partial x_0$,
$\partial H_0/\partial p_0=\partial H_1/\partial p_1$, and 
$\partial H_0/\partial p_1=-\partial H_1/\partial p_0$, 
the equations of motion reduce to 
\begin{eqnarray}
{\dot p}_0=-\frac{\partial H_0}{\partial x_0}, \quad 
{\dot x}_0=\frac{\partial H_0}{\partial p_0}, \quad 
{\dot p}_1=\frac{\partial H_0}{\partial x_1}, \quad {\rm and} \quad 
{\dot x}_1=-\frac{\partial H_0}{\partial p_1} . 
\label{eq:x12}
\end{eqnarray}
In particular, since the total energy $H_0+i H_1$ is conserved under 
the Hamiltonian dynamics, if the initial energy is real so that $H_1=0$, 
then it will remain real. Thus, a complexified Hamiltonian system is 
equivalent to a real two-dimensional system for which the position 
and momentum are interchanged in the second dimension 
\cite{Xavi,kaushal}. 

The investigations of `complexified' quantum mechanics, on the other hand, 
consider complex Hamiltonians of the form $\hat H=\hat H_0+i \hat H_1$, 
while typically leaving the quantum phase space real. The equation of 
motion is still given by the Schr\"odinger equation. That is, writing 
$z_n=(x_n+i p_n)/\sqrt{2}$ for the components of the state vector, we have 
$i \dot z_n= \rmd H/\rmd {\bar z}_n$, where the Hamiltonian function is 
the expectation value of $\hat H$, which is now complex. It is then 
straightforward to verify that equations of motion for the real phase 
space variables $p_n,\, x_n$ are given by
\begin{eqnarray}
{\dot p}_n=-\frac{\partial H_0}{\partial x_n} - 
\frac{\partial H_1}{\partial p_n}  \quad {\rm and} \quad 
{\dot x_n}=\frac{\partial H_0}{\partial p_n}-
\frac{\partial H_1}{\partial x_n} .
\label{eq:xx13}
\end{eqnarray}
This structure is a combination of a Hamiltonian 
symplectic flow generated by $H_0$ and a Hamiltonian 
gradient flow generated by $H_1$, and also appears 
in the semiclassical limit of non-Hermitian quantum dynamics 
\cite{Graefe}. 

The dynamics of complexified quantum systems 
considered in the literature, for the most part, are characterised by 
(\ref{eq:xx13}), and not by (\ref{eq:x12}). This is the sense in which 
the nature of complexification in classical mechanics has been 
different from that of quantum mechanics. To see in which way 
these two formalisms might be unified by a quaternionic or 
coquaternionic approach, we shall now investigate the special 
case of a bilinear Hamiltonian, before turning to the more general 
case.

\textit{Quaternionic and coquaternionic oscillators}. 
We again examine a one-dimensional dynamical system 
with the Hamiltonian $H={\bar z}z$, but now $z$ is assumed a 
quaternion. That is, although the dynamical variable is given by 
$z = (x + i p)/\sqrt{2}$, we let $p$ and $x$ extend into the 
complex domain according to the prescription $p\to p_0+j p_1$ 
and $x \to x_0 + j x_1$. It follows from the algebraic property 
(\ref{eq:2}) that $z=(x_0 + i p_0 + j x_1 + k p_1)/\sqrt{2}$. 
We assume for simplicity a superselection rule \cite{Finkelstein} 
that singles out the complex number $i$ for determining the 
direction of time. Hence the canonical equation of motion is 
given by (\ref{eq:7}) with $H=p_0^2+p_1^2+x_0^2+x_1^2$, and 
we obtain 
\begin{eqnarray}
\left\{ \begin{array}{ll} 
{\dot p}_0=-x_0 & \ {\dot p}_1=-x_1 \\ 
{\dot x}_0=p_0 & \ {\dot x}_1=p_1 . 
\end{array} \right. \label{eq:10}
\end{eqnarray}
Thus, a quaternionic oscillator is equivalent to a two-dimensional 
oscillator. We remark that working with the real variables $(p,x)$, 
a natural way of extending the theory into quaternionic domain 
would not have been apparent. 

Let us compare this system with a complex PT-symmetric oscillator 
\cite{Bender2}, where we take the Hamiltonian $H=\frac{1}{2}(p^2+x^2)$ 
for which equations of motion read ${\dot p}=-x$ and ${\dot x}=p$, 
and extend the phase space variables into the complex domain: 
$p\to p_0+i p_1$ and $x \to x_0 + i x_1$. The resulting dynamical 
equations are given by (\ref{eq:10}). While equations of motion 
agree, and indeed they share the same fixed-point structure, 
these two theories are nevertheless distinct because of 
boundary conditions: The energy $E=p_0^2+p_1^2+x_0^2+x_1^2$ 
is positive definite for quaternionic oscillators; whereas the energy 
$E=p_0^2-p_1^2+x_0^2-x_1^2$  is indefinite for PT-symmetric 
oscillators. 

Next we consider a coquaternionic oscillator, with the same 
superselection rule. (It is worth remarking that unlike quaternionic 
quantum mechanics where one has to make the choice for the 
direction of time, in coquaternionic quantum mechanics the 
imaginary unit $i$ is naturally preferred on account of the fact 
that $i^2=-1$, while $j^2=k^2=+1$.) 
In this case, equations of motion remain 
the same, but owing to the split-signature of coquaternions the 
energy is given by $E=p_0^2-p_1^2+x_0^2-x_1^2$. Hence we 
conclude that a complex PT-symmetric oscillator is equivalent to 
a coquaternionic oscillator. An important point to note here is the 
fact that in the complex formulation of mechanics the construction 
of the Hamiltonian such as $H={\bar z}z$ implicitly involves the 
notion of an inner product, which is not evident in Hamilton's 
formulation involving real variables $(p,x)$. In the case of a complex 
PT-symmetric oscillator, 
this inner product is in effect determined by a PT conjugation, which, 
owing to the fact that the parity operator is trace free, has a split 
signature equivalent to a coquaternionic norm. 

At the level of classical mechanics, the appearance of a PT inner 
product seems unproblematic because it has no impact on equations 
of motion. However, if we attempt to formulate statistical mechanics, 
for instance, then the use of a PT inner product raises a severe 
obstacle: For any bounded energy, the energy shell in phase space 
is not compact, hence an equilibrium microcanonical distribution 
cannot be defined. At the quantum level, the issue of indefiniteness 
in a PT inner product has been recognised earlier because inner 
products are related to probabilities, which have to be nonnegative. 
To remedy this issue, an alternative inner product, based on a CPT 
conjugation, has been introduced \cite{BBJ,Mostafa}. One possible 
way forward in the case of a complex classical oscillator therefore is 
to modify the inner product such that indefinite components pick up 
an additional minus one. Then energy shells for finite energies 
become compact, allowing, in particular, for a rigorous formulation of 
statistical mechanics.   

\textit{Coquaternionic mechanics with real and complex energies}. 
More generally, consider now a real-valued Hamiltonian $H(z,{\bar z})$ 
where $z=(x_0 + i p_0 + j x_1 + k p_1)/\sqrt{2}$ is a coquaternion, and 
hence $\sqrt{2}\,\partial/\partial{\bar z} = \partial/\partial x_0 - i 
\partial/\partial p_0 + j \partial/\partial x_1 + k \partial/\partial p_1$. From 
(\ref{eq:7}) we thus deduce that 
\begin{eqnarray}
i{\dot z} =  \left[ \frac{\partial H}{\partial x_0} + i \frac{\partial H}{\partial 
p_0} - j \frac{\partial H}{\partial x_1} - k \frac{\partial H}{\partial p_1} \right], 
\label{eq:x10}
\end{eqnarray}
and hence that
\begin{eqnarray}
{\dot p}_0=-\frac{\partial H}{\partial x_0}, \quad 
{\dot x}_0=\frac{\partial H}{\partial p_0}, \quad 
{\dot p}_1=\frac{\partial H}{\partial x_1}, \quad {\rm and} \quad 
{\dot x}_1=-\frac{\partial H}{\partial p_1} . 
\label{eq:x11}
\end{eqnarray}
Comparing these equations with (\ref{eq:x12}) where the initial 
energy is assumed real, we find therefore that a 
PT-symmetric classical system with real energy is equivalent 
to a coquaternionic mechanical system with real energy. Note 
that in a quaternionic mechanical system the role of $p_1$ and 
$x_1$ are interchanged, and thus it is merely equivalent to a 
two-dimensional real system. 

An alternative case to consider is where the phase space or the 
state space variables are kept real so that $z$ remains a complex 
number, but the energy $H=H_0+i H_1$ is made complex. In this 
case, it follows from (\ref{eq:7}) that equations of motion read 
\begin{eqnarray}
{\dot p}=-\frac{\partial H_0}{\partial x} + 
\frac{\partial H_1}{\partial p}  \quad {\rm and} \quad 
{\dot x}=\frac{\partial H_0}{\partial p} + 
\frac{\partial H_1}{\partial x} , 
\label{eq:x13}
\end{eqnarray}
which is of course a special case of (\ref{eq:xx13}). If we 
now allow both energy and phase space variables be complex, then 
in the coquaternionic case, we obtain:
\begin{eqnarray} 
&& 
{\dot p}_0=-\frac{\partial H_0}{\partial x_0}+\frac{\partial H_1}{\partial p_0}, 
\quad 
{\dot x}_0=\frac{\partial H_0}{\partial p_0}+\frac{\partial H_1}{\partial x_0}, 
\nonumber \\ && 
{\dot p}_1=\frac{\partial H_0}{\partial x_1}+\frac{\partial H_1}{\partial p_1}, 
\quad  
{\dot x}_1=-\frac{\partial H_0}{\partial p_1}+\frac{\partial H_1}{\partial x_1} . 
\label{eq:x14}
\end{eqnarray}
These equations constitute natural generalisations of (\ref{eq:x13}) in 
the sense that they embody the structure of a combination of a 
Hamiltonian symplectic flow generated by $H_0$ and a 
Hamiltonian gradient flow generated by $H_1$. 

We remark, more generally, that if the Hamiltonian is coquaternionic-valued 
so that $H=H_0+iH_1+jH_2+kH_3$, then (\ref{eq:x10}) implies 
\begin{eqnarray} 
&& \fl
{\dot p}_0=-\frac{\partial H_0}{\partial x_0}+\frac{\partial H_1}{\partial p_0}
+\frac{\partial H_2}{\partial x_1}+\frac{\partial H_3}{\partial p_1}, 
\quad 
{\dot x}_0=\frac{\partial H_0}{\partial p_0}+\frac{\partial H_1}{\partial x_0}
-\frac{\partial H_2}{\partial p_1}+\frac{\partial H_3}{\partial x_1}, 
\nonumber \\ && \fl
{\dot p}_1=\frac{\partial H_0}{\partial x_1}+\frac{\partial H_1}{\partial p_1}
-\frac{\partial H_2}{\partial x_0}+\frac{\partial H_3}{\partial p_0}, 
\quad  
{\dot x}_1=-\frac{\partial H_0}{\partial p_1}+\frac{\partial H_1}{\partial x_1} 
+\frac{\partial H_2}{\partial p_0}+\frac{\partial H_3}{\partial x_0}
\label{eq:x14a}
\end{eqnarray}
for the dynamical equations. 

\textit{Two oscillators and two-level systems}. 
The orbit space of a classical system of a pair of oscillators is 
equivalent mathematically (though not physically) to a quantum 
two-level system. In particular, the reduced phase space of both 
systems is just the Bloch sphere $S^2$. This can be seen by 
noting that the Hamiltonian of a pair of uncoupled oscillators in 
complex coordinates is $H={\bar z}_1z_1+{\bar z}_2z_2$. If the 
energy is fixed we obtain the three-sphere ${\bar z}_1z_1+
{\bar z}_2z_2=1$, upon which the Hamiltonian flow acts as scalar 
multiplication $z_n\to\re^{i t}z_n$ \cite{Atiyah}. Hence the orbit 
space is just the two-sphere resulting from the Hopf map 
$S^3 \to S^2$. 

Similarly, a pair of classical quaternionic oscillators is 
equivalent to a quaternionic two-level quantum system. In this 
case, once energy is fixed we obtain the seven-sphere ${\bar z}_1
z_1+{\bar z}_2z_2=1$, upon which the Hamiltonian flow acts as 
quaternionic scalar multiplication. Hence the associated phase 
space (Bloch sphere) is a four-sphere $S^4$ resulting from the 
Hopf map $S^7 \to S^4$. To see the structure of the state 
space in the quaternionic case more explicitly, it suffices to note that 
a generic normalised state vector in a quaternionic two-level 
system can be written in the parametric form 
\begin{eqnarray}
 \left( \begin{array}{l} z_1 \\ z_2 \end{array} \right) 
 = \left( \begin{array}{l} \cos\frac{1}{2} \theta \\ 
\sin\frac{1}{2} \theta\, \re^{{\boldsymbol i}_\phi \phi_1} 
\end{array} \right), 
\end{eqnarray} 
where 
${\boldsymbol i}_\phi = i \cos\phi_2 + j \sin\phi_2 \cos\phi_3 
+ k \sin\phi_2 \sin\phi_3$. It is then straightforward to see that 
$(\theta,\phi_1,\phi_2, \phi_3)$ constitute spherical coordinates 
for an $S^4$ in ${\mathds R}^5$. In 
higher dimensions, the state space is just the quaternionic 
projective Hilbert space endowed with a quaternionic 
Fubini-Study metric arising from transition probabilities. 
In the case of a pair of coquaternionic oscillators, the resulting 
orbit space is hyperbolic (a four-dimensional analogue of the 
Poincar\'e disk, which is an example of a Siegel domain 
\cite{Vinberg}). 

In the case of a quaternionic or coquaternionic two-level system, a 
generic Hermitian Hamiltonian can be expressed in the form 
\begin{eqnarray}
{\hat H} = \half \omega {\mathds 1} + \sum_{l=1}^5 n_l {\hat\sigma}_l, 
\label{eq:13}
\end{eqnarray}
where $\omega\in{\mathds R}$, ${\vec n}$ is a unit vector on 
$S^4\subset{\mathds R}^5$, and 
\begin{eqnarray}
{\hat\sigma}_1 &=& \left( \begin{array}{cc} 0 & 1 \\ 
1 & 0 \end{array} \right), \quad\! 
{\hat\sigma}_2 = \left( \begin{array}{cc} 0 & -i \\ 
i & 0 \end{array} \right), \quad\!
{\hat\sigma}_3 = \left( \begin{array}{cc} 1 & 0 \\ 
0 & -1 \end{array} \right), \nonumber \\ && 
{\hat\sigma}_4 = \left( \begin{array}{cc} 0 & -j \\ 
j & 0 \end{array} \right), \quad\! 
{\hat\sigma}_5 = \left( \begin{array}{cc} 0 & -k \\ 
k & 0 \end{array} \right) \label{eq:17}
\end{eqnarray} 
are the (co)quaternionic Pauli matrices. For a quaternionic 
Hermitian matrix the eigenvalues are real; whereas they 
are either real or appear as complex conjugate pairs in the case 
of a coquaternionic Hermitian matrix. In either case, for 
two-level systems there are six parametric degrees of freedom 
(consistent with the observation that the most general 
PT-symmetric $2\times2$ Hamiltonian has six exogenous parameters 
\cite{wang}). 
In standard quantum mechanics, it follows from the construction of 
spin-orbit interaction (the so-called Pauli correspondence) that one can 
identify the spin operators with the three spatial directions. Similarly, 
(\ref{eq:17}) suggests that in a quaternionic or coquaternionic quantum 
mechanics, the ambient space has five dimensions. 

It is worth remarking that owing to the lack of commutativity 
of the imaginary units there are ``observable'' effects in quaternionic 
or coquaternionic mechanics that have no analogue in real mechanics. 
For instance, a Hamiltonian proportional 
to an identity matrix can generate nontrivial dynamics \cite{Wolff}. 
In particular, if we let $i$ determine the preferred complex 
structure, and take the Hamiltonian ${\hat H}=\frac{1}{2}\omega 
{\mathds 1}$, then the dynamics gives rise to a Rabi oscillation 
between the ${\hat\sigma}_4$ and ${\hat\sigma}_5$ directions. 
This can be seen most easily in the Heisenberg picture: 
\begin{eqnarray}
{\hat\sigma}_4(t) = \cos(\omega t) {\hat\sigma}_4 + 
\sin(\omega t) {\hat\sigma}_5,
\end{eqnarray}
where we have chosen the initial state to be the ${\hat\sigma}_4$ 
eigenstate. 
Therefore, we can generate dynamics with vanishing energy 
gap. This is reminiscent of the so-called ``arbitrary fast'' quantum 
state transport effect \cite{BBJM,Mostafa2,Gunther}. 

Let us now consider a $2\times2$ coquaternionic Hermitian 
Hamiltonian and express this in the form 
\begin{eqnarray}
{\hat H}=& \left( \begin{array}{cc} s+t & q \\ 
{\bar q} & s-t \end{array} \right), 
\end{eqnarray}  
where $q=q_0+iq_1+jq_2+kq_3$. The eigenvalues are 
$E_{\pm} = s \pm \sqrt{t^2 + {\bar q}q}$,  
where ${\bar q}q = q_0^2 + q_1^2 - q_2^2 - q_3^2$. Therefore, in 
the coquaternionic representation, if $t^2+q_0^2 + q_1^2 > q_2^2 + 
q_3^2$, then the eigenvalues are real, which can be identified as the 
region of unbroken PT symmetry; otherwise, the eigenvalues form a 
complex conjugate pair. 

Working in 
the basis for which ${\hat H}$ is diagonal, the phase-space 
function for the Hamiltonian is $H={\bar z}_1 E_+ z_1 + 
{\bar z}_2 E_- z_2$ (note that in general $E_\pm$ and $z_i$ 
need not commute, since $E_\pm$ need not be real). It follows 
that equations of motion are given by ${\dot z}_{1}=i E_{+} z_{1}$ 
and ${\dot z}_{2}=i E_{-} z_{2}$. In particular, if 
PT symmetry is unbroken, then eigenvalues are real and 
the fixed points associated with the dynamical evolution are 
centres; whereas if the symmetry is broken, then the fixed 
points form a vortex pair. This is a characteristic feature of 
the so-called PT phase transition whereby a pair of centres 
coalesce and turn into a pair of vortices (a sink and a source) 
\cite{Graefe3}. Such a transition does not occur in the case 
of a quaternionic quantum theory, and this establishes the 
inequivalence of PT-symmetric and quaternionic quantum 
theories. 

\textit{Discussion}. 
We have shown that PT-symmetric quantum theories are 
equivalent to certain coquaternionic extensions of Hermitian 
quantum theories, at least in finite-dimensional cases. 
This equivalence suggests that the underlying space-time dimension 
is six, not four, on account of the two additional dimensions 
generated by the additional imaginary units. This follows from 
the fact that spin-orbit interaction can only be defined for five spatial 
dimensions in quaternionic or coquaternionic quantum theories. 

Needless to say, there is a substantial literature dealing with various 
aspects of six-dimensional space-times. Apart from the practicality 
that calculus on a six-dimensional space facilitates various analysis 
in four dimensions \cite{dirac,hughston,weinberg2}, models for 
six-dimensional space-times have also been proposed at a fundamental 
level. Empirical evidences for extra dimensions are often sought at 
inaccessibly high energies. However, by identifying physical phenomena 
predicted by coquaternionic quantum theories that cannot be described 
by standard complex quantum mechanics (such as the Rabi oscillation 
with vanishing energy gap described above), it is possible to seek 
compelling evidences for the existence of extra dimensions at low 
energies (cf. \cite{Adler3} for a discussion on difficulties in detecting 
quaternionic manifestations in scattering experiments). To this end, it is 
worth remarking that evidences for octonionic quantum mechanics 
would suggest that the underlying space-time has dimension ten.  

Some of the challenging open issues of PT-symmetric quantum 
theories include the formulation of combined systems to characterise 
such notions as entanglement and coherent states. The mathematical 
difficulty might be rooted in the fact that quaternionic projective Hilbert 
spaces (unlike complex projective space of ordinary quantum 
mechanics \cite{Brody}) do not admit algebraic varieties that characterise 
subspaces of particular physical characteristics \cite{Berndt}. 
Nevertheless, progress has been made in the context of 
quaternionic quantum mechanics to formulate combined systems 
\cite{Horwitz} or coherent states \cite{Adler2,THK}. It may be that 
these techniques can be applied to investigate open issues associated 
with combined systems for PT-symmetric quantum theories. 

\vskip 6pt
EMG is supported by an Imperial College Junior Research Fellowship.
\vskip 10pt 



\end{document}